\documentclass[twocolumn,aps,prl,nofootinbib,superscriptaddress, tightenlines]{revtex4}

\usepackage{amsmath, amsfonts, amsthm, amssymb, graphicx, color,hyperref}
\usepackage{subfigure}

\allowdisplaybreaks[3]

\newcommand{\bwt} {\begin{widetext}}
\newcommand{\ewt} {\end{widetext}}
\newcommand{\be} {\begin{equation}}
\newcommand{\ee} {\end{equation}}
\def \bal#1\eal  {\begin{align} #1 \end{align}}

\newcommand{\pd} {\partial}
\newcommand{\bfx} {{\bf x}}

\newcommand{\mc} {\mathcal}

\begin{document}

\title{Charge-Swapping Q-balls}

\author{Edmund J.~Copeland}
\author{Paul M.~Saffin}
\affiliation{School of Physics and Astronomy, University of Nottingham, Nottingham, NG7 2RD, UK}

\author{Shuang-Yong Zhou}
\affiliation{SISSA, Via Bonomea 265, 34136, Trieste, Italy and INFN, Sezione di Trieste, Italy}
\affiliation{Department of Physics, Case Western Reserve University, 10900 Euclid Ave, Cleveland, OH 44106, USA}

\date{\today}

\begin{abstract}

Q-balls are non-topological solitonic solutions to a wide class of field theories that possess global symmetries. Here we show that in these same theories there also exists a tower of novel composite Q-ball solutions where, within one composite Q-ball, positive and negative charges co-exist and swap at a frequency lower than the natural frequency of an individual Q-ball. These charge-swapping Q-balls are constructed by assembling Q-balls and anti-Q-balls tightly such that their nonlinear cores overlap. 
We explain why charge-swapping Q-balls can form and why they swap charges.

\end{abstract}

\maketitle

Many field theories have non-linear, extended, stable solutions, known as solitons. Some solitons are called ``topological defects'', as their existence is implied by nontrivial topological mappings between the coordinate space and the field space; for excellent reviews see \cite{Manton:2004tk,Vachaspati:2006zz}. Q-balls belong to another class of soliton that are time-dependent but non-dissipative, whose stability is guaranteed by the existence of Noether charges associated with some continuous global symmetries \cite{Friedberg:1976me,Coleman:1985ki}. 

Consider the simplest case of a complex scalar field $\Phi$ with U(1)-symmetric Lagrangian
\be
\mc{L} = -\pd_\mu \Phi \pd^\mu \bar{\Phi} - V(\phi)   ~,
\ee
where $\Phi=(\phi_1+i\phi_2)/\sqrt{2}$ and $\phi=  \sqrt{\phi_1^2+\phi_2^2}$, $\phi_1$ and $\phi_2$ being real fields. A particle of this field has a conserved charge, denoted as $Q$, associated with the $U(1)$ symmetry. A Q-ball is a lump of coherent particles, i.e., a lump of $Q$ charges, whose classical configuration is described by $\Phi=\varphi(r)e^{i\omega t}$, where $r$ is the radius from the center of the ball, $\varphi(r)$ is a real monotonically decreasing function and $\omega$ is the real constant angular velocity in the internal field space. Similarly, an anti-Q-ball is a lump of coherent anti-particles with negative charges. Coleman has shown that isolated Q-balls exist and are a preferred, stable configuration for a large class of potentials that ``open up'' away from their minimum  \cite{Coleman:1985ki}. Assuming, without loss of generality, that the potential has its minimum at $V(0)=0$ and the perturbative mass squared around the vacuum, $V''(0)$, is positive, the ``open-up'' condition requires that the minimum of $2V(\phi)/\phi^2$ be at some $\phi$ away from $0$ \cite{Coleman:1985ki}. This condition can be interpreted as the energy per particle in a Q-ball being smaller than the perturbative mass, thus implying that a Q-ball is stable against particle emission.

Q-balls are found in several areas of physics, for example, phase transitions in the early universe. They exist in supersymmetric extensions of the Standard Model, where there are typically plenty of flat directions in the scalar potential where the baryon or lepton number is conserved, and Affleck-Dine condensates can form along these directions \cite{Affleck:1984fy, Dine:1995uk}, which can then fragment into Q-balls \cite{Kusenko:1997si} (see \cite{Dine:2003ax, Enqvist:2003gh} for a review). The fragmentation can in turn produce a stochastic gravitational wave background \cite{Kusenko:2008zm} with a characteristic multi-peak structure in the power spectrum \cite{zhou}. Q-balls may account for dark matter \cite{Kusenko:1997si, Dine:2003ax, Enqvist:2003gh}, which makes up one quarter of the energy budget in the current universe. Condensed matter systems in the laboratory can also give rise to Q-balls \cite{Enqvist:2003zb}. An explicit example has been realized experimentally by Bunkov and Volovik in superfluid ${}^3$He-B, in which the coherent precession of magnetization of the superfluid is described by a magnon condensate and the order parameter of the condensate plays the role of the $U(1)$ scalar field \cite{volovik}.

While the single Q-ball solution has the simple form $\Phi=\varphi(r)e^{i\omega t}$ with $\varphi(r)$ simply obtained from solving an ODE, multi-Q-ball solutions, or Q-ball interactions, are generally rather complicated. Some interesting phenomena, such as the phase dependence of forces between Q-balls, charge transfer during collisions and Q-ball fission and fusion, have been revealed \cite{Axenides:1999hs,Battye:2000qj,Bowcock:2008dn}, but overall this is a largely under-explored area.

\begin{center}
\begin{figure}[ht]
\centering
\subfigure[]{%
\includegraphics[height=0.75in,width=0.75in]{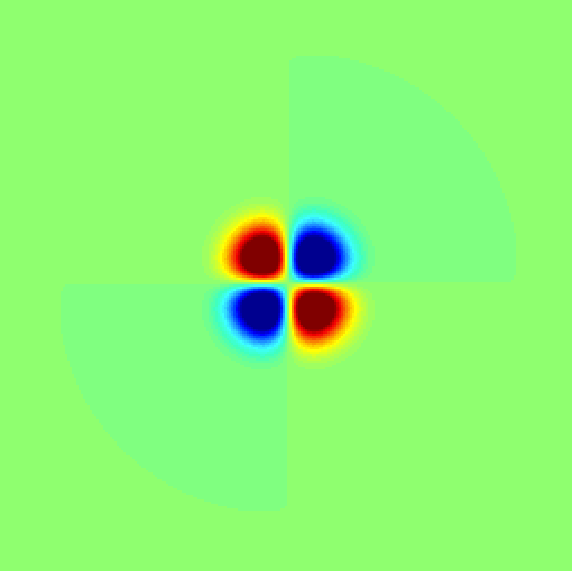}
}
\subfigure[]{%
\includegraphics[height=0.75in,width=0.75in]{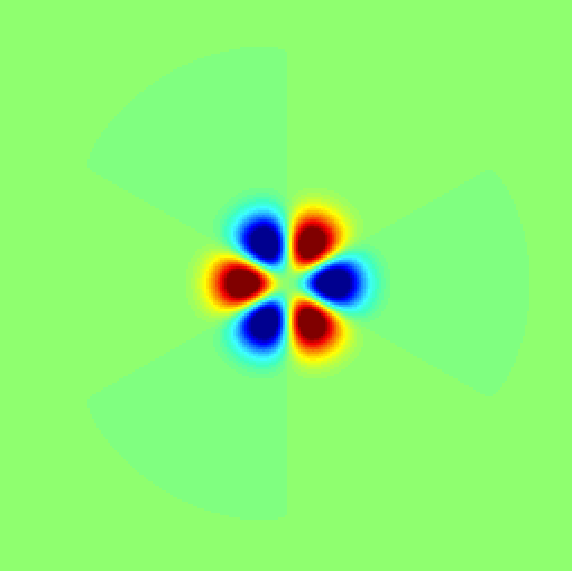}
}
\subfigure[]{%
\includegraphics[height=0.75in,width=0.75in]{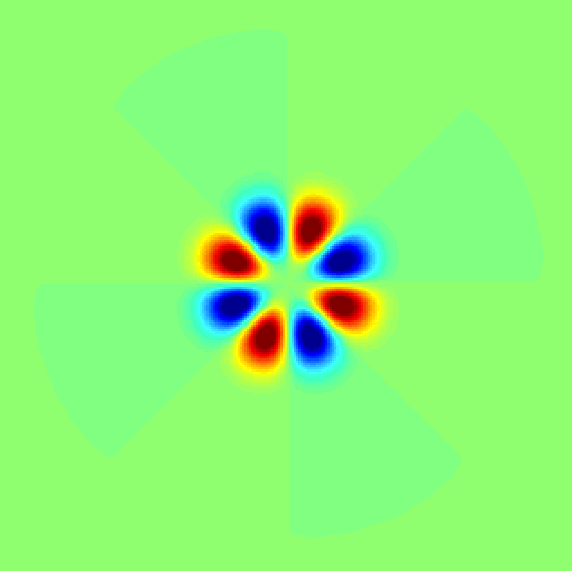}
}

\subfigure[]{%
\includegraphics[height=0.75in,width=0.75in]{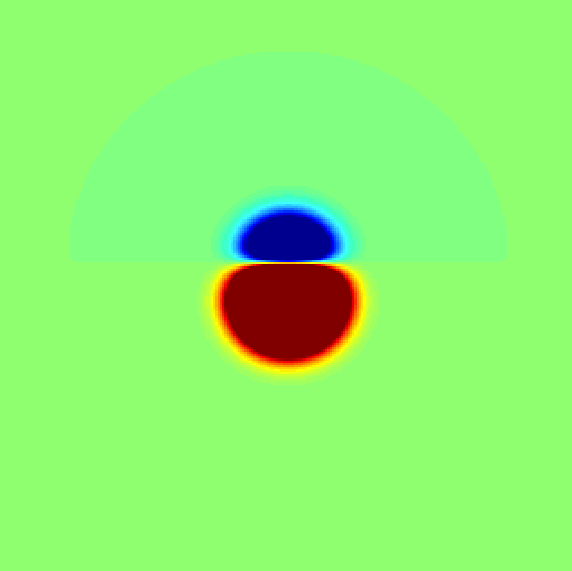}
}
\subfigure[]{%
\includegraphics[height=0.75in,width=0.75in]{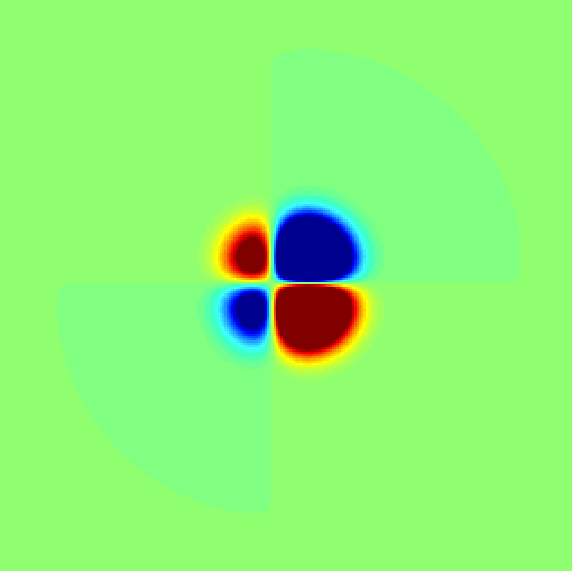}
}
\subfigure[]{%
\includegraphics[height=0.75in,width=0.75in]{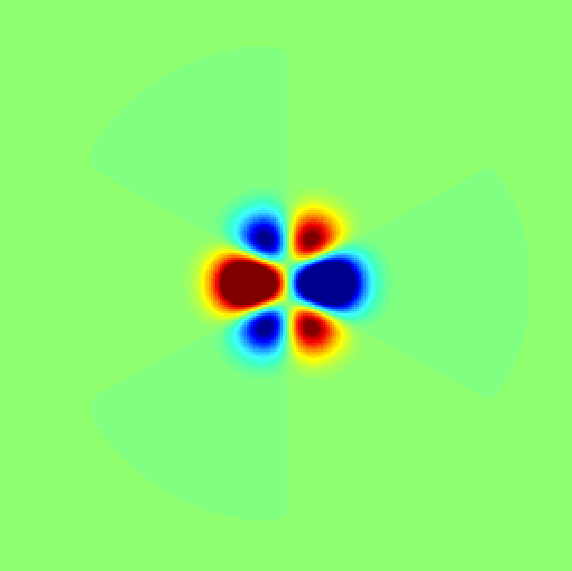}
}

\caption{A few examples of initial CSQ configurations with more Q-balls and anti-Q-balls and/or with unequal charges. They evolve in a manner similar to the Q-ball anti-Q-ball case, with the positive and negative charges swapping place as the system evolves.}
\label{fig:hoCSQ}
\end{figure}
\end{center}

In this Letter, we describe a new exciting phenomenon in Q-ball interactions: A Q-ball and an anti-Q-ball can form a bound state that has the fascinating property that the opposite charges of the two Q-balls swap at a frequency lower than the natural oscillation frequency of each constituent Q-ball. We refer to these new objects as charge-swapping Q-balls (CSQs). Furthermore, we can build a whole tower of higher rank CSQs with more Q-balls and anti-Q-balls -- see Fig.~\ref{fig:hoCSQ} for a few simple examples. Thus, in this way,  we can construct a new class of non-linear localized solutions in a field theory where single Q-ball solutions exist.

Although we have constructed CSQs for other potentials, the numerical results in this Letter are presented for the running mass potential
\be
\label{logpotential}
V(\phi) =  \frac12 m^2\phi^2 \left[1+ K \ln \left(\frac{\phi^2}{2M^2}\right)\right] ~,
\ee
with $K=-0.1$ and $M=10m$ as the fiducial parameters. This is a typical effective potential taking into account the renormalization effects on the soft mass term in supersymmetric models \cite{Enqvist:1997si,Enqvist:2003gh}, so the Q-balls in this potential may have a physical bearing in early universe phase transitions. Although this potential is unbounded from below when $\phi$ is very large, this is not a concern for us, as our initial conditions are such that the only part of the potential that is accessed is the region bounded-from-below. For the realistic supersymmetric flat direction potential, there is usually an extra term $\phi^{2n}$ with $n$ being some positive integer, which stabilizes the potential against quantum tunneling. We have compared our simulation results both with and without the $\phi^{2n}$ term included, and, as expected, find little difference. 

Another interesting feature of this potential, is that it is a separable potential that admits a Gaussian profile for a single Q-ball, which has led to its appearance in many previous studies \cite{bbm,Bogolyubsky:1976yu,Bogolyubsky:1976nx,Ventura:1976va,Ferraz de Camargo:1982sk,Garriga:1994ut}. For our parameters, the single Q-ball solution is given by 
$\Phi(t,\bfx) = A T(t)X(\bfx)$, where $T(t)=e^{i\omega t}$, $X(\bfx)= e^{-r^2/2\sigma^2}$, $\sigma=1/\sqrt{-Km^2}$ and $A=M e^{(\omega^2-m^2+2m^2K)/2m^2K}$. Although very useful, we should emphasize that the existence of CSQs does not rely on this separable property associated to this potential. 

To prepare CSQs, we simply superimpose single Q-balls and anti-Q-balls sufficiently close to each other as the initial configuration, i.e. for $n$ Q-balls located at ${\bf d}_n$ with angular velocity $\omega_n$ and phase $\alpha_n$ we have,
\be
\label{CSQsup}
\Phi_{\rm initial}  = \sum_n    
A_n e^{\frac{(\bfx-{\bf d}_n)^2}{2\sigma^2}} e^{i\omega_n t + \alpha_n}  ~,
\ee
where $A_n=M e^{(\omega_n^2-m^2)/2m^2K+1}=10m \,e^{-5\omega_n^2+6}$ and $\sigma=1/\sqrt{-Km^2}\simeq 3.16/m$ for our fiducial model. The initial configuration relaxes to the CSQ solution very quickly. Here we restrict ourselves to the simpler case of vanishing initial relative velocities. There are physical reasons to do this as it mimics a number of realistic scenarios such as when Q-balls are formed during a phase transition.  It is of course also possible to choose the parameters $A_n$ and $\sigma$ to be different from the above values, which would correspond to choosing excited Q-balls in the initial state.

\begin{center}
\begin{figure}[ht]
\centering
\subfigure[~$t=0/m$]{%
\includegraphics[height=0.75in,width=0.75in]{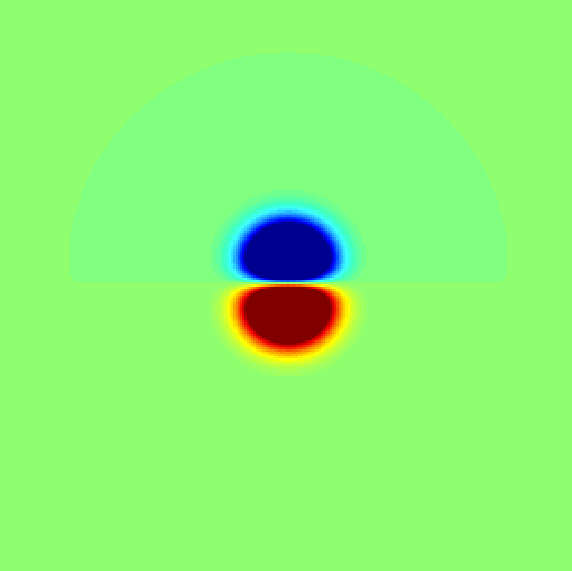}
}
\subfigure[~$t=12/m$]{%
\includegraphics[height=0.75in,width=0.75in]{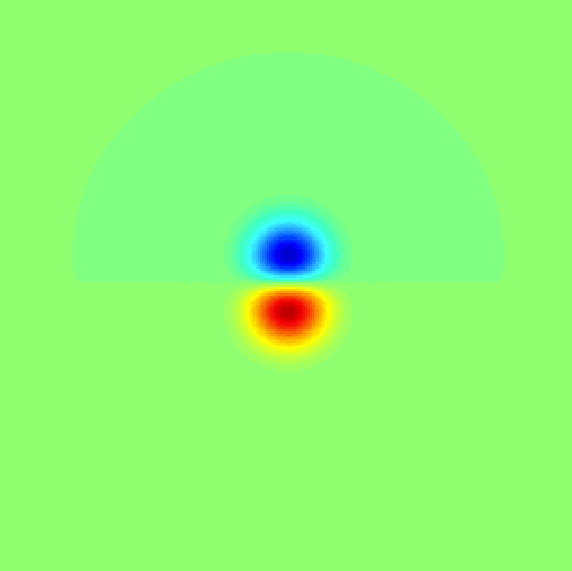}
}
\subfigure[~$t=17/m$]{%
\includegraphics[height=0.75in,width=0.75in]{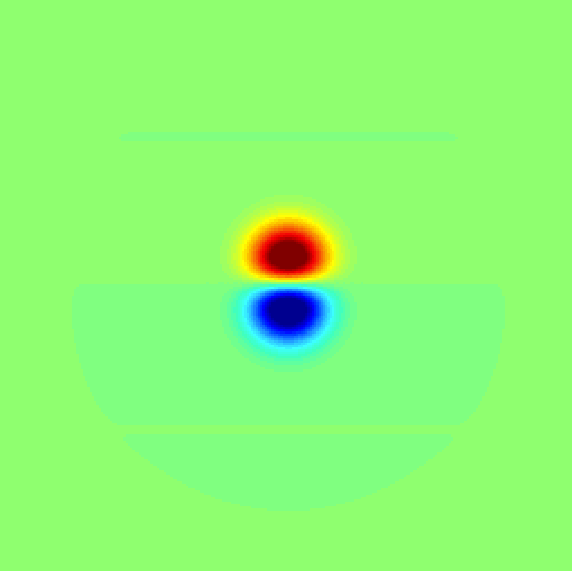}
}
\subfigure[~$t=29/m$]{%
\includegraphics[height=0.75in,width=0.75in]{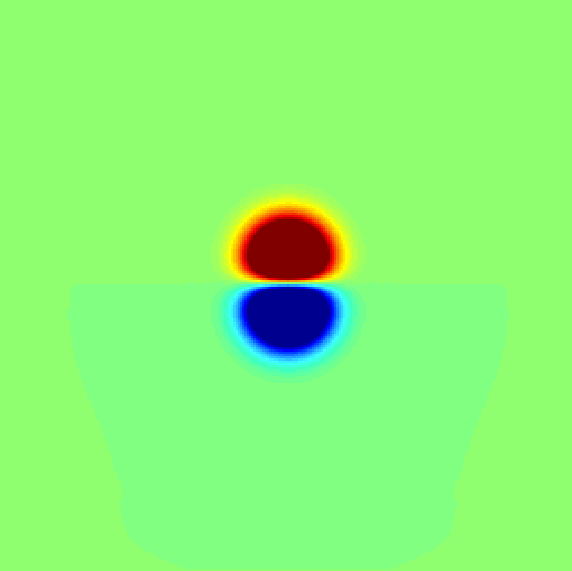}
}

\subfigure[~$t=40/m$]{%
\includegraphics[height=0.75in,width=0.75in]{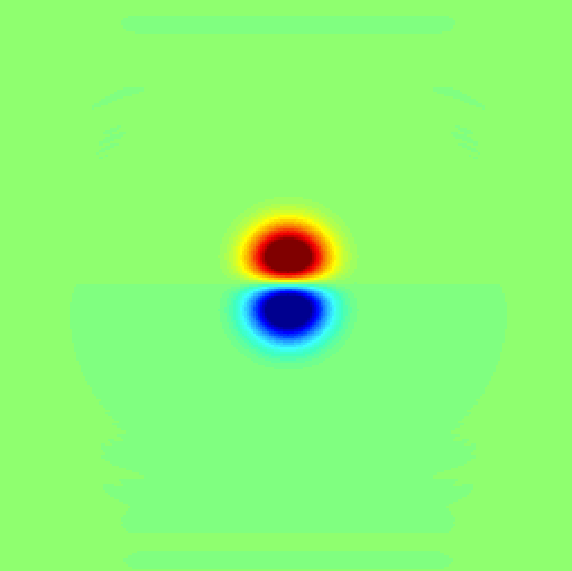}
}
\subfigure[~$t=45/m$]{%
\includegraphics[height=0.75in,width=0.75in]{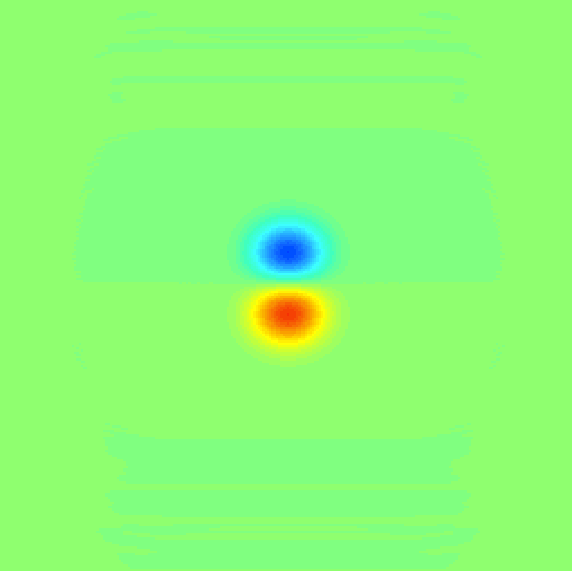}
}
\subfigure[~$t=56/m$]{%
\includegraphics[height=0.75in,width=0.75in]{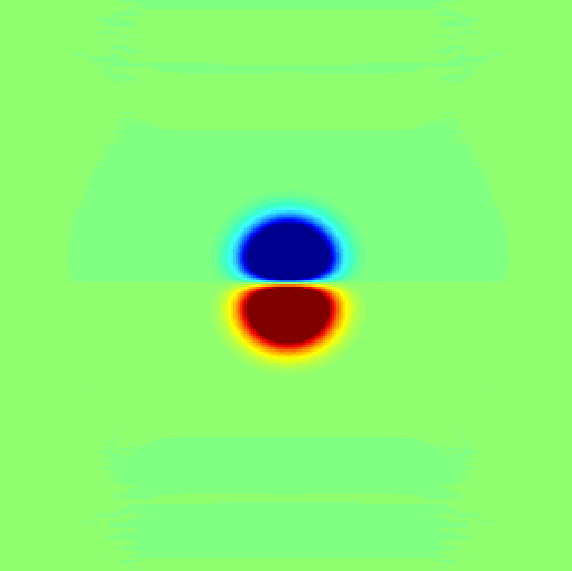}
}
\subfigure[energy dens.]{%
\includegraphics[height=0.75in,width=0.75in]{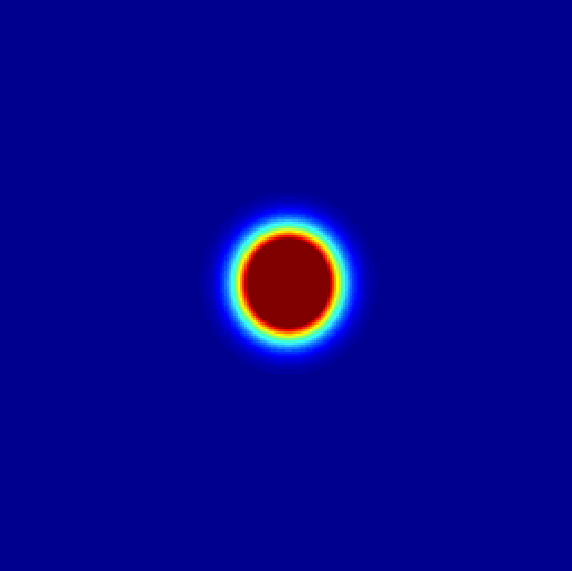}
}
\caption{Evolution of the charge density ($(a)$--$(g)$) and energy density ($(h)$) for a charge-swapping Q-ball prepared from a Q-ball and an anti-Q-ball with equal and opposite charges. The energy density profile is almost unchanged throughout the evolution. For $(a)$--$(g)$, the color scheme of blue to red ranges from $-20m^3$ to $20m^3$, while for $(h)$ it is from $0$ to $100m^4$.}
\label{fig:twoBallEvo}
\end{figure}
\end{center}

We first consider the simplest case of a Q-ball and an anti-Q-ball with equal and opposite charges: $\omega_1=-1.1m,~\omega_2=1.1m$, $\alpha_1=\alpha_2=0$, $|{\bf d}_1 - {\bf d}_2| = 1.6/m $. In this case, we have a composite Q-ball where the positive and negative charges swap with a frequency lower than the natural angular frequency (the latter being roughly $m$). Meanwhile the energy density remains like that of a single Q-ball. See Fig.~\ref{fig:twoBallEvo} for a full period of charge swapping in 2+1 dimensions. The CSQ lives for at least $\mc{O}(10^4)$ natural oscillation periods, as long as our simulations reliably run. Note that Coleman's Q-ball stability theorem \cite{Coleman:1985ki}, which states that the single Q-ball solution minimizes the energy functional, is applicable to a system with a sufficiently large total charge. Physically, this is because when a large number of the charged particles group together as a Q-ball, the system's total energy is lowered. For our case, the total charge is zero, so Coleman's stability theorem does not apply. We can, however, say that these CSQs cannot be absolutely stable, as we can smoothly deform the CSQ configuration to an outward propagating wave with arbitrarily small amplitudes while keeping the total energy fixed and the total charge zero. Nevertheless, they may be stable to small perturbations, which would be consistent with our simulations.

\vskip -12pt
The existence of CSQs depends on the separation of the two constituent Q-balls $|{\bf d}_1 - {\bf d}_2|$.  Empirically, CSQs can form when the constituent Q-balls are sufficiently close to each other:
\be
|{\bf d}_1 - {\bf d}_2| \lesssim 2\sigma  ~.
\ee
The physical meaning of this empirical condition is very clear: One may regard the inner $\sigma$ region of the single Q-ball as the core, and CSQs exist when there is overlap between the cores of the two Q-balls, i.e., when nonlinearity becomes important. On the other hand, as we explain shortly, we find that the existence of CSQs does not depend on the initial relative phase, i.e. $\alpha_1-\alpha_2 $ can be arbitrary.

\begin{center}
\begin{figure}[ht]
\centering
\includegraphics[height=1.5in,width=2.6in]{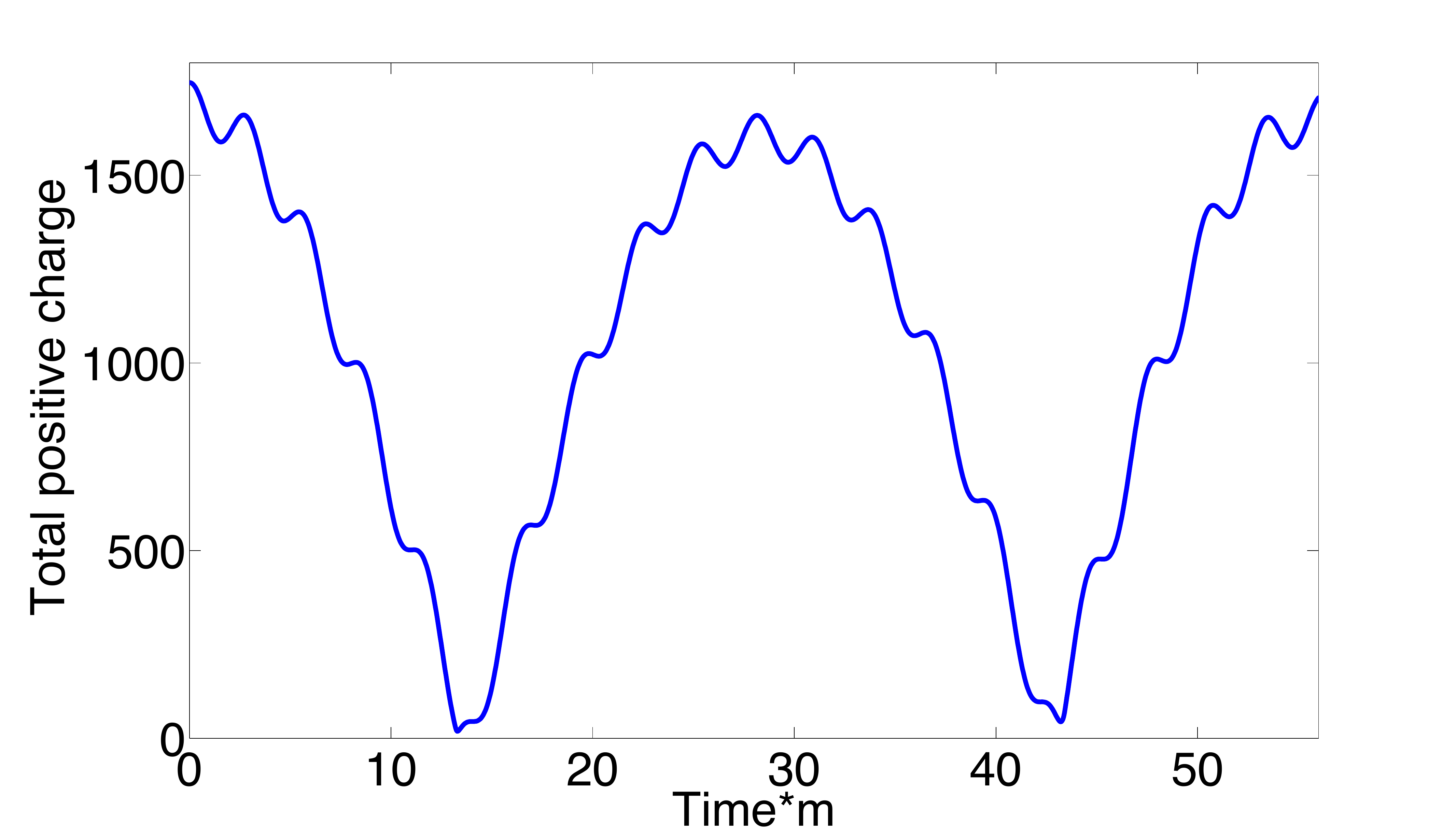}
\caption{The evolution of the sum of positive charges over the same period as Fig.~\ref{fig:twoBallEvo}.} 
\label{fig:plotPC}
\end{figure}
\end{center}

To qualitatively understand the charge swapping phenomenon in terms of particle interactions, we first note that the energy per particle in a Q-ball is less than the perturbative mass \cite{Coleman:1985ki}, so a Q-ball is a very dense concentration of  coupled particles. Also, while the channel of positive and negative charges annihilating into gauge bosons is absent for an ungauged field theory, there are processes, such as the $2\Phi+2\bar{\Phi}\to \Phi + \bar{\Phi}$ process ($\Phi$ indicating a particle and $\bar{\Phi}$ an anti-particle), that can reduce the positive and negative charges separately. When the cores of a Q-ball and an anti-Q-ball meet, due to high density, the dominating processes are those charge reducing interactions such as $2\Phi+2\bar{\Phi}\to \Phi + \bar{\Phi}$. As the charges reduce, the Q-ball and anti-Q-ball appear to disappear. But, when the particle density is sufficiently lowered and the remaining particles become sufficiently relativistic, the charge increasing processes such as $\Phi+\bar{\Phi}\to 2\Phi + 2\bar{\Phi}$ begin to become dominant and the Q-ball and anti-Q-ball appear to re-emerge. This constitutes half of the charge swapping period of Fig.~\ref{fig:twoBallEvo}. Fig.~\ref{fig:plotPC} backs up the above argument as it describes the evolution of the sum of all the positive charges for the same period as that of Fig.~\ref{fig:twoBallEvo}. Also, the current conservation equation is given by $\dot{\rho}_Q = \pd_i ( \phi_1\pd_i{\phi}_2-\pd_i{\phi}_1\phi_2)$, where $\rho_Q=\phi_1\dot{\phi}_2-\dot{\phi}_1\phi_2$. From this, we can see that any change in the charge density is balanced by the spatial flux of the charge ``momenta'', which explains the spatial movement of charges.

\begin{center}
\begin{figure}[ht]
\centering
\includegraphics[height=1.6in,width=2.13in]{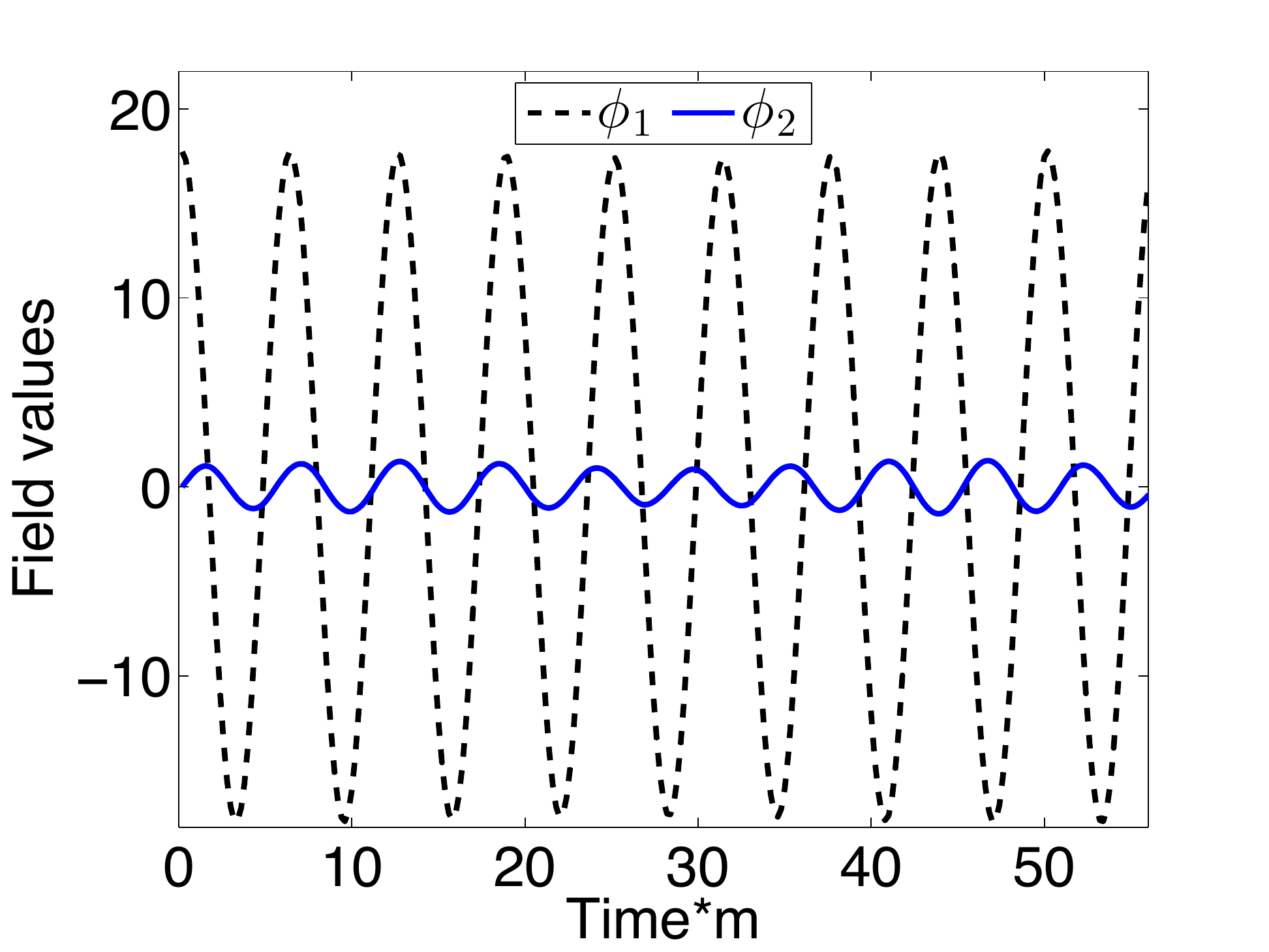}
\caption{The evolution of $\phi_1$ and $\phi_2$ for a point close to the border between the positive and negative charges over the same period as Fig.~\ref{fig:twoBallEvo}.} 
\label{fig:phi1phi2}
\end{figure}
\end{center}

\vskip -12pt
Quantitatively, we may understand both why the charges swap and why the swapping frequency is lower than the natural oscillation frequency as follows. Given the form of the charge density and the oscillating nature of $\phi_1$ and $\phi_2$, the sign of the charge density is determined by the relative phase between $\phi_1$ and $\phi_2$. In Fig.~\ref{fig:phi1phi2}, we plot the evolution of $\phi_1$ and $\phi_2$ for a point close to the border between the positive and negative charges for the same period as Fig.~\ref{fig:twoBallEvo}. We can see $\phi_2$ oscillates at a frequency slightly higher than that of $\phi_1$, which slowly changes the phase difference with time, resulting in a lower charge swapping frequency. For example, the charge density $\rho_Q$ at the chosen point is positive initially, but when $t\simeq 13/m$ and $t\simeq 44/m$, both $\phi_1$ and $\phi_2$ reach their extreme amplitudes corresponding to $\dot{\phi_1}=\dot{\phi_2} \simeq 0$ hence to 
$\rho_Q$ almost vanishing, which matches with the positive charge evolution in Fig.~\ref{fig:plotPC}. The reason why $\phi_2$'s frequency is higher than that of $\phi_1$ is linked to the fact that the equations of motion for both $\phi_1$ and $\phi_2$ are nonlinear and that they have different amplitudes. For our fiducial model, the logarithmic interaction in Eq (\ref{logpotential}) does not make the $\phi_1$ and $\phi_2$'s oscillations deviate significantly away from the corresponding harmonic cases. Instead, it effectively shifts the oscillating frequency by a small amount. As $K$ is negative, a small field amplitude (i.e. $\phi_2$ here) gives rise to a higher effective frequency.

A useful way of understanding why a Q-ball and an anti-Q-ball can attract to form a composite Q-ball is by considering a Q-ball as two interacting oscillons, which are composed of different fields, $\phi_1$ and $\phi_2$. An oscillon \cite{Bogolyubsky:1976yu,Gleiser:1993pt,Copeland:1995fq} is also a localized, oscillating configuration of a nonlinear field, and the existence of oscillon solutions requires the potential has an ``open-up'' feature, analogous to a Q-ball. But unlike a Q-ball, there is no stabilizing Noether charge associated with an oscillon. So an oscillon by itself is long-lived, simply due to the non-linear nature of the field theory, but will ultimately decay. In the Q-ball context, the $\phi_1$ oscillon and the $\phi_2$ oscillon are linked via appropriate interactions such that both of them are stabilized. As the field space of an oscillon is one dimensional, there is a clear picture of whether two oscillons of the same field attract or repel each other when their initial separation is small ($\lesssim 2\sigma$ for a Gaussian profile oscillon). Indeed, we find that there exists a surprisingly simple relation: For two oscillons with approximately equal amplitudes, they attract each other when they are close to being in phase (their phase difference $\lesssim \pi/2$) and they repel each other when roughly in anti-phase (their phase difference $\gtrsim \pi/2$).

\begin{center}
\begin{figure}[ht]
\centering
\subfigure[]{%
\includegraphics[height=1.2in,width=1.2in]{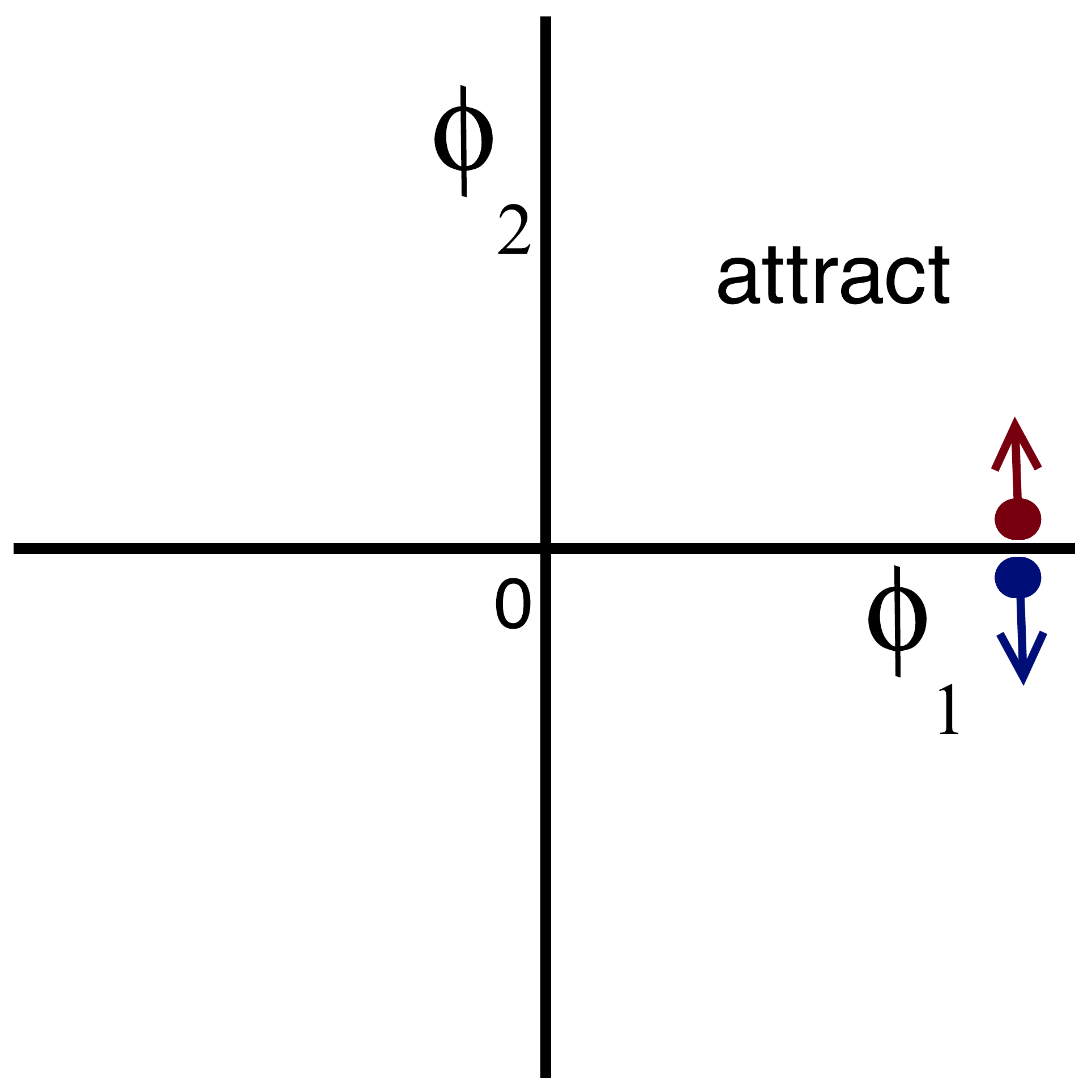}
}
\subfigure[]{%
\includegraphics[height=1.2in,width=1.2in]{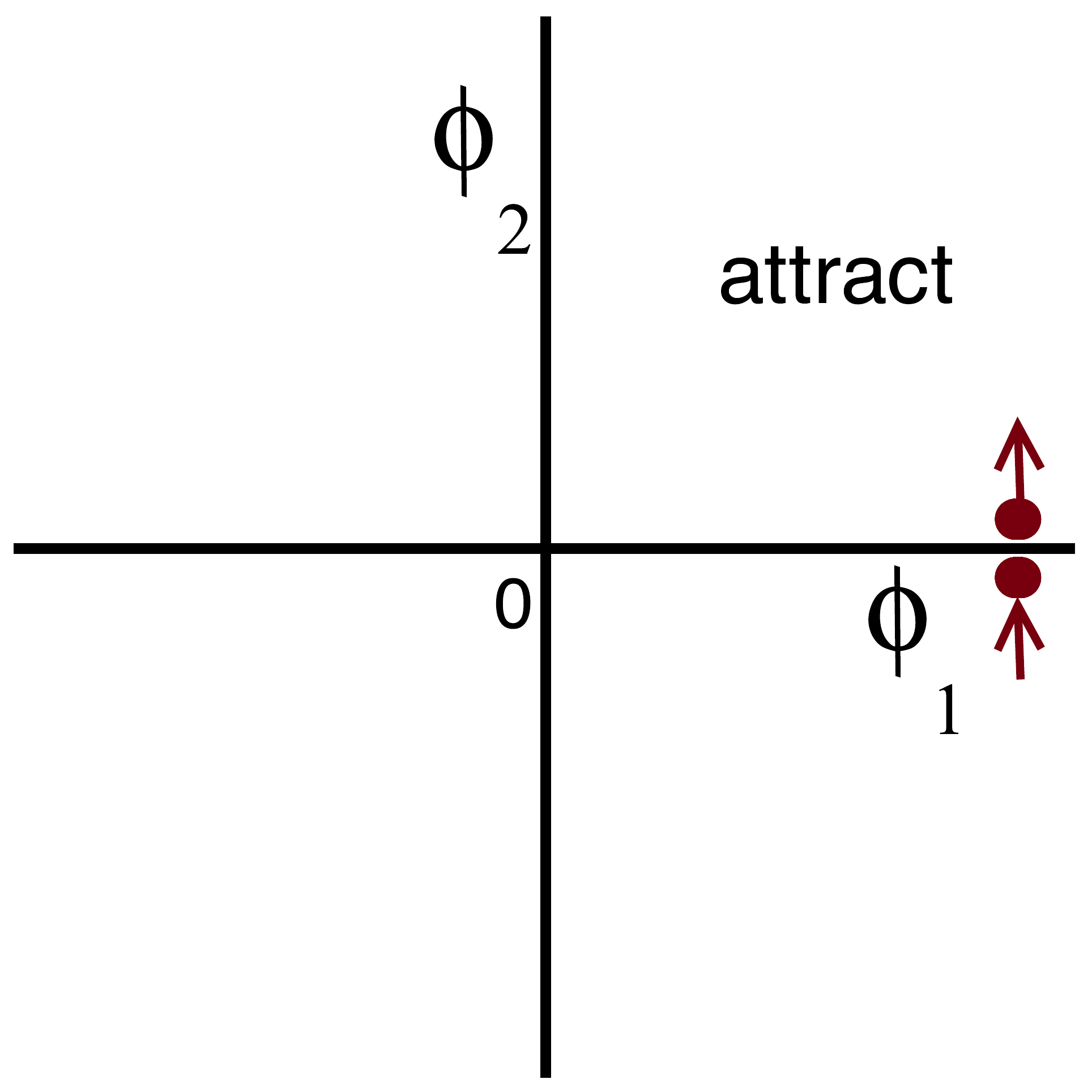}
}
\subfigure[]{%
\includegraphics[height=1.2in,width=1.2in]{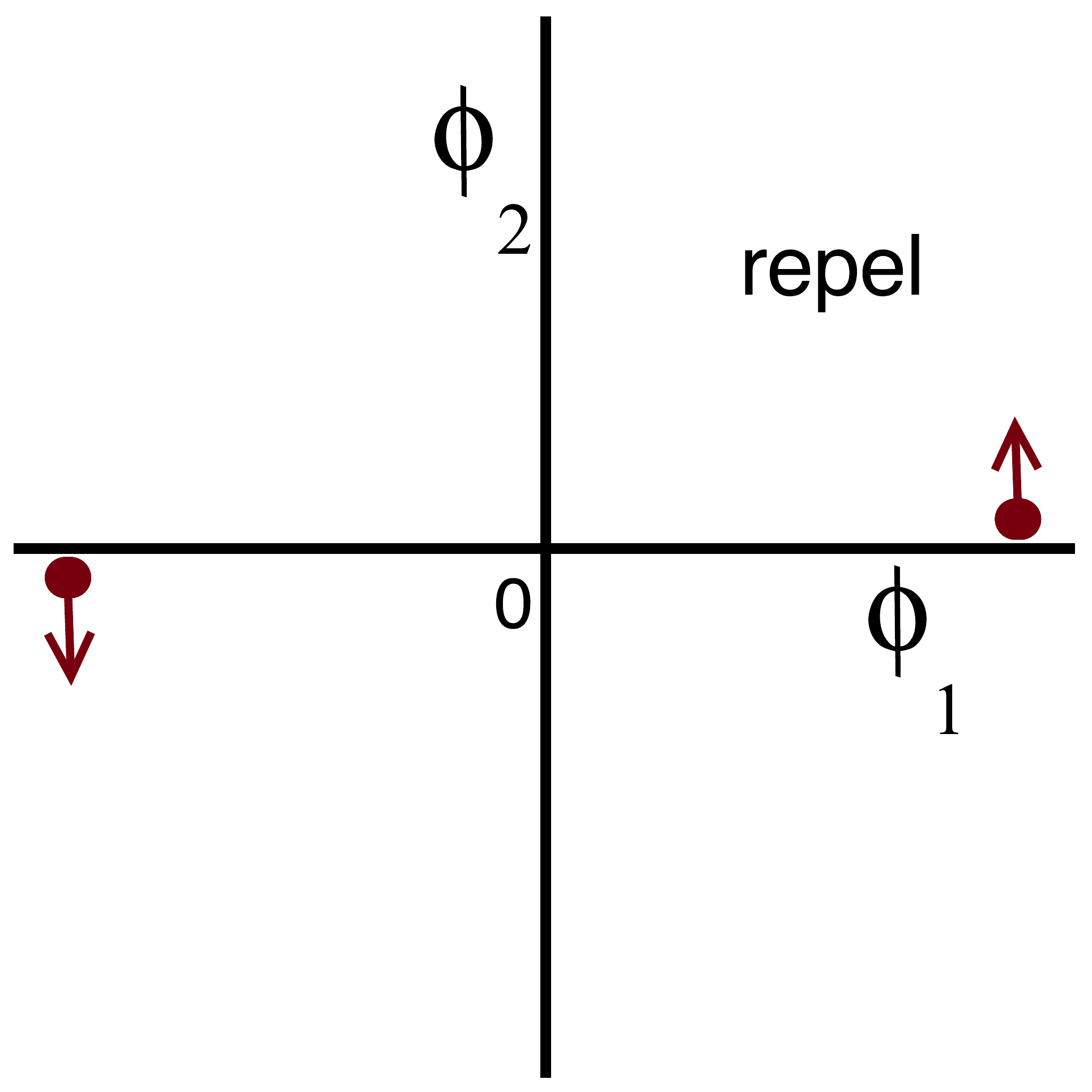}
}
\subfigure[]{%
\includegraphics[height=1.2in,width=1.2in]{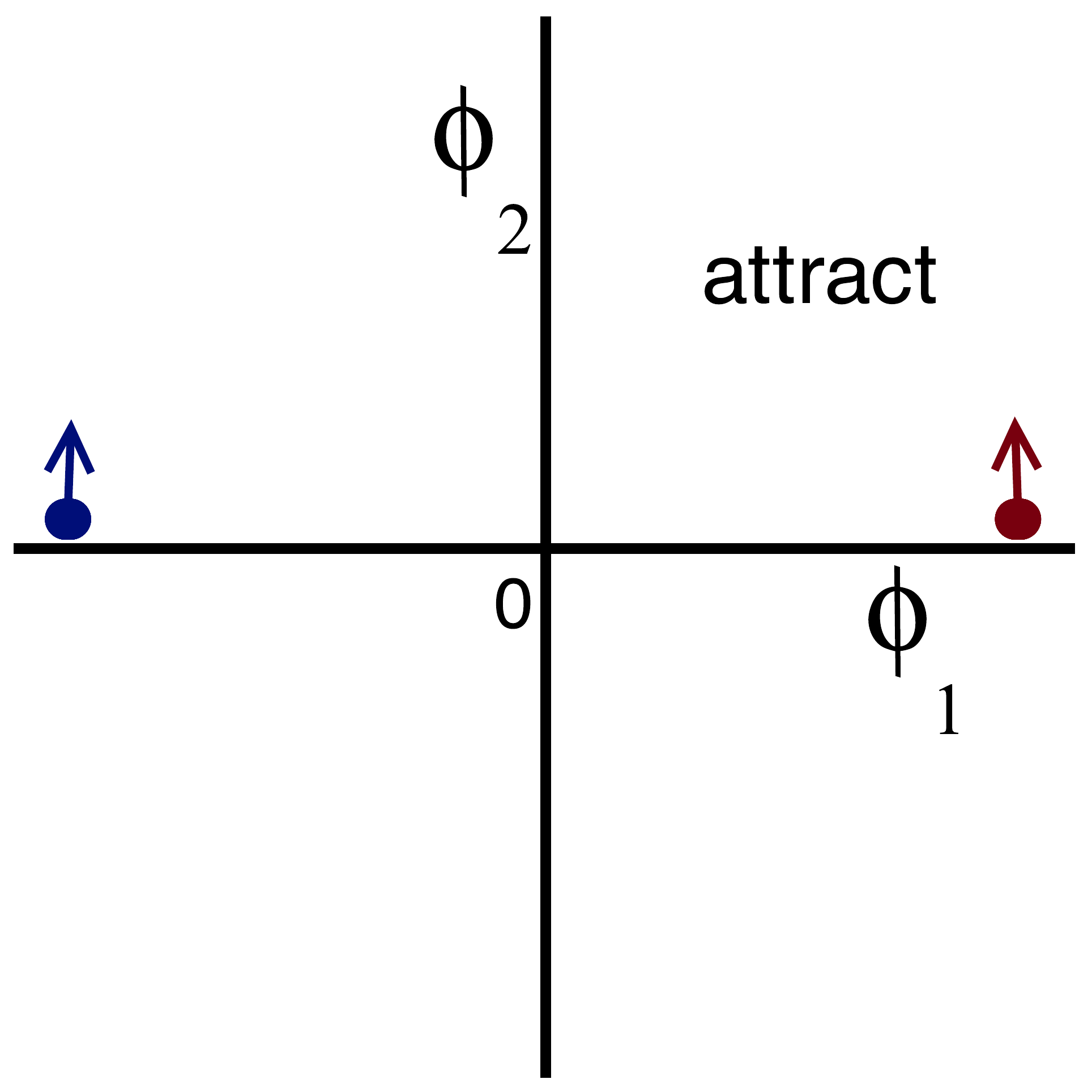}
}
\caption{Initial ($t=0$) phase alignments of two closely placed Q-balls and whether they attract or repel. The vertical (horizontal) axis corresponds to $\phi_2$ ($\phi_1$). A red solid circle represents a Q-ball's position in the internal field space of $\Phi$, and a blue solid circle presents that of an anti-Q-ball. The arrows represent the velocity directions (sign of $\omega$) of the corresponding Q-balls in field space. $(a)$ attract: the $\phi_1$ fields of a Q-ball and anti-Q-ball are in phase (i.e. they track each other in the $\phi_1$ direction) and the $\phi_2$ fields are in anti-phase (i.e. they head off in opposite directions in the $\phi_2$ direction). $(b)$ attract: $\phi_1$, $\phi_2$ both in phase. $(c)$ repel: $\phi_1$, $\phi_2$ both in anti-phase. $(d)$ attract: $\phi_1$ in anti-phase and $\phi_2$ in phase.
}
\label{fig:phases}
\end{figure}
\end{center}

Now, if we view a Q-ball roughly as a $\phi_1$ oscillon plus a $\phi_2$ oscillon, there are four cases of extreme initial phase alignments, depending on whether the $\phi_1$ oscillon and the $\phi_2$ oscillon are in phase or in anti-phase. Fig.~\ref{fig:phases} presents the four initial ($t=0$) phase alignments  in the internal field space of $\Phi$ (the vertical axis corresponding to $\phi_2$ and the horizontal one to $\phi_1$) and whether these alignments correspond to attractive or repulsive configurations. While case (b) of Fig.~\ref{fig:phases}  is a single charge bound state, cases $(a)$ and $(d)$ correspond to CSQs.  We can see that as long as there is one kind of oscillon-pair that is in phase, either the $\phi_1$ or $\phi_2$ kind, the two Q-balls attract, despite the fact that the other kind of oscillons may be in anti-phase. This suggests that in this two Q-ball setup the attracting force from two in-phase oscillons is generically greater than the repelling force from two in-anti-phase oscillons. The reason for this is simple: the $\phi_1$ and $\phi_2$ oscillons have equal amplitudes when the Q-balls are well separated, but when they are placed close together, on average the amplitudes for the in-phase oscillons are enhanced while the amplitudes for the in-anti-phase oscillons are reduced (See Fig.~\ref{fig:phi1phi2}), leading to a net attraction.

\vskip -12pt
With the phase alignment argument, it is also easy to see that the conditions for the existence of CSQs is insensitive to the initial phase difference of the two constituent Q-balls. For case $(b)$ and $(c)$ of Fig.~\ref{fig:phases}, the phase alignment will be roughly maintained during the early evolution, since the two Q-balls have the same kind of charge so that they rotate in the same direction in field space. For case $(a)$ and $(d)$ of Fig.~\ref{fig:phases}, however, the Q-ball and the anti-Q-ball are rotating in the opposite direction, so case $(d)$ becomes case $(a)$ following roughly a quarter of a period of evolution - up to a rotation in field space. In fact, for a Q-ball and an anti-Q-ball with a generic initial phase difference, its phase alignment will always become case $(a)$ after some fraction of a period of evolution - up to a rotation in field space. Therefore the initial phase difference plays little role in forming CSQs.

Finally, we have been mainly focusing on CSQs constructed from a Q-ball and an anti-Q-ball with equal and opposite charges. But, as we have mentioned, we can also construct CSQ configurations with more Q-balls and anti-Q-balls and moreover with unequal charges. Apart from the CSQ of Fig.~\ref{fig:twoBallEvo}, the simplest members in this new class of nonlinear objects are shown in Fig.~\ref{fig:hoCSQ}, where each of the configurations leads to long-lived charge-swapping solitons. Hints of more complicated, chaotic CSQs for the running mass potential have appeared in simulations with random initial conditions \cite{Enqvist:2000cq,Hiramatsu:2010dx}, but it was not known whether they have CSQs' intrinsic properties such as the charge swapping and the energy density resembling that of the single Q-ball. We have reproduced these results and found they indeed possess the CSQ properties.

\vskip -12pt
In conclusion, we have described a new composite state of non-topological solitons, the charge-swapping Q-ball (CSQ). The positive and negative charges within one CSQ swap with a frequency lower than the natural oscillating frequency of the constituent Q-balls, whilst the energy density of the CSQ resembles that of a single Q-ball. Their existence can be understood in terms of forces between multiple oscillons with different phases. Although we have not presented details here, we have obtained CSQs in various dimensions and models. We have shown that the CSQs can be obtained through a relaxation of appropriately superimposed single Q-ball and anti-Q-ball solutions, thus suggesting a practical way to construct a new series of highly non-trivial, localized nonlinear solutions for field theories that possess single Q-ball solutions. As single Q-balls have been constructed in laboratories, we speculate that the novel properties of CSQs may also find their applications in condensed matter systems.

\vskip 5pt
\noindent{\bf Acknowledgment}\\
We would like to thank Kari Enqvist, Anupam Mazumdar, Mitsuo Tsumagari and Jian Zhou for helpful discussions.  EJC and PMS are grateful to the STFC for financial support. SYZ acknowledges support from DOE grant DE-SC0010600.

\end{document}